\documentclass[aps,prb,twocolumn,superscriptaddress]{revtex4}
\usepackage{amssymb}
\usepackage{graphicx}
\usepackage{amsmath}
\usepackage{bm}

\begin{document}

\title{Unsupervised Learning of Frustrated Classical Spin Models I: Principle Component Analysis}

\author{Ce Wang}
\affiliation{Institute for Advanced Study, Tsinghua University, Beijing, 100084, China}
\author{Hui Zhai}
\affiliation{Institute for Advanced Study, Tsinghua University, Beijing, 100084, China}
\affiliation{Collaborative Innovation Center of Quantum Matter, Beijing, 100084, China}

\date{\today }

\begin{abstract}

This work aims at the goal whether the artificial intelligence can recognize phase transition without the prior human knowledge. If this becomes successful, it can be applied to, for instance, analyze data from quantum simulation of unsolved physical models. Toward this goal, we first need to apply the machine learning algorithm to well-understood models and see whether the outputs are consistent with our prior knowledge, which serves as the benchmark of this approach. In this work, we feed the compute with data generated by the classical Monte Carlo simulation for the $XY$ model in frustrated triangular and union jack lattices, which has two order parameters and exhibits two phase transitions. We show that the outputs of the principle component analysis agree very well with our understanding of different orders in different phases, and the temperature dependences of the major components detect the nature and the locations of the phase transitions. Our work offers promise for using machine learning techniques to study sophisticated statistical models, and our results can be further improved by using principle component analysis with kernel tricks and the neural network method. 

\end{abstract}

\maketitle

\section{Introduction}

The most ambitious goal for applying machine learning to physics research is to discover new physics without prior human knowledge. For an example, in the near future one practical use is to analysis data from quantum simulation. The ultimate goal of quantum simulation is to perform simulating quantum computation on physical models that can not be solved by a classical computer. That is to say, there is no reliable prior human knowledge on these models. One of such example is the quantum simulation of the Fermi Hubbard model \cite{Hubbard}. The Fermi Hubbard model lies at the heart of understanding many strongly correlated materials, however, it can not be solved by reliable classical computational methods once the filling is away from half-filling, and the exact diagnoalization is limited to very small size, because of which it has been an issue of debate for several centuries. Recently, experiments on ultracold Fermi atoms in optical lattices can now simulate the Fermi Hubbard model with a system size much larger that the computation capability of a classical computer \cite{FH_exp}. The output of these quantum simulation experiments will be a huge number of data measured at different temperature and different parameters. Hence, it becomes a crucially important problem that how one can analyze these big data to extract useful, and some times latent, physical information, without prior knowledge of the model. From this perspective, artificial intelligence could be useful.   

On the way toward this goal, we should first see that whether the machine learning algorithm can reproduce the known results of well-understood models, which serves as an important benchmark for the machine learning approach. For this purpose, we focus on the classical statistical models. For these models, the classical Monte Carlo simulation can generate lots of configurations at different temperature and different parameter regimes, and they are used as input data to feed the computer. Over the past many decades, physicists have developed the concept of ``order parameter", with which these configurations can be classified into different phases; while computer does not have the concept of order parameter as a prior knowledge. Nevertheless, through the algorithm developed in machine learning studies\cite{PRML}, computer can also successfully group the configurations into different classes. We will show that the classification by machine learning algorithm is consistent with the order parameter classification of these models. Further, by adding the information of temperature associated with each configuration, computer can also determine the critical temperature of the phase transition. 

\begin{figure}[b]
\includegraphics[width=3.0 in]
{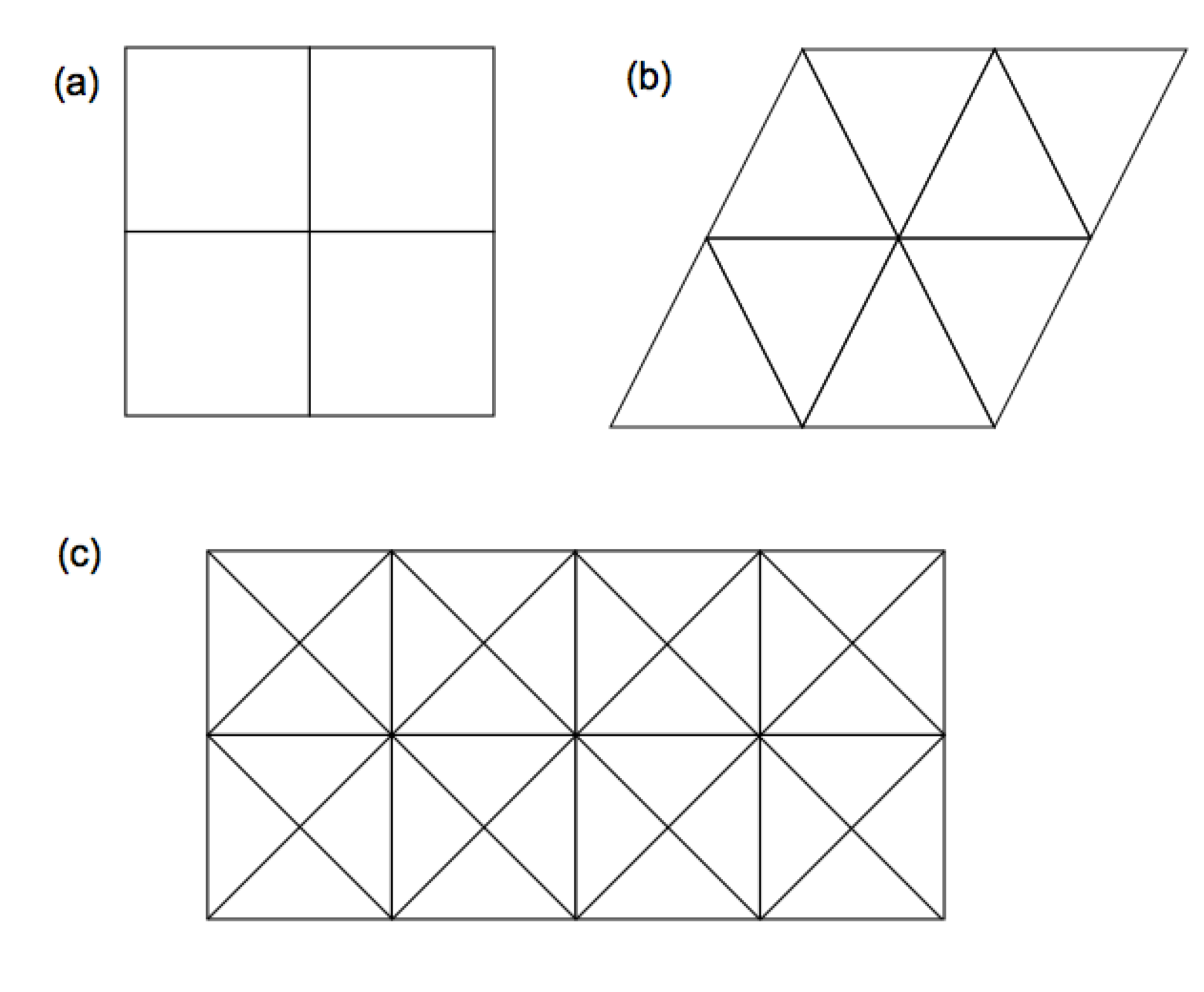}
\caption{Schematic of (a) square, (b) triangular and (c) union jack lattices.}
\label{lattice}
\end{figure}

The recent studies along this line have investigated the Ising model and some other related models, by applying various machine learning methods such as the principle component analysis with or without kernel tricks and the artificial neural network\cite{Ising_PCA,Ising_nn,Confusion,Ising_BM,Ising_XY_VAE,Ising_SVM,SqXY_PCA,Ising_SU(2)_nn}. The classical $XY$ model with continuous variables on an unfrustrated square lattice has also been studied\cite{Ising_XY_VAE,SqXY_PCA}. Some works have also studied quantum models of strongly correlated fermions and accelerated Monte Carlo simulations\cite{CNN_Fermions1,CNN_Fermions2,Liang3,Liang2,Liang1,WangLei1,WangLei2,WangLei3}. In this work we study the classical $XY$ model in frustrated lattices\cite{FXY_theory5,FXY_theory6,FXY_theory1,FXY_theory2,FXY_theory3,FXY_theory4,FXY_theory0,Monte_tri_uj,Monte_tri}, including both triangular and union jack lattices. The key advance is that these models are frustrated and exhibit two order parameters, that are, a $U(1)$ order parameter and a chirality order parameter, and consequently, two successive phase transitions as lowering the temperature\cite{FXY_theory0,Monte_tri_uj,Monte_tri}. We show that the machine learning method can successfully capture both. Our work represents an important step toward applying machine learning to sophisticated physical models.

\section{Model and Method}

In this work we consider the $XY$ model whose Hamiltonian is given by 
\begin{equation}
\mathcal{H}=J\sum\limits_{\langle ij\rangle}\cos(\theta_i-\theta_j),
\end{equation}
where $\theta_{i}\in(0,2\pi]$ is a classical variable defined at each site, $\langle ij\rangle$ denotes all the nearest neighboring bonds. Below we will consider the lattice structures including two-dimensional square, triangular and union jack lattices, respectively, as shown in Fig. \ref{lattice}. We use classical Monte Carlo to generate equilibrium configurations, denoted by $\{x_n\}$ ($n=1,\dots,N$), at a lattice with totally $L$ lattice sites and at different temperatures. $N$ is the total number of data set, and $N=N_T\times N_0$, where $N_0$ is the number of configurations generated at each given temperature, and $N_T$ is the number of different temperatures considered. At each site-$i$, the spin is described by its orientation $(\cos\theta_i,\sin\theta_i)$, and in this case, each $x_n$ is a vector of $2L$-dimension, and for later convenience, they are organized as 
\begin{equation}
x_n=(\cos\theta_1, \dots,\cos\theta_L,\sin\theta_1,\dots,\sin\theta_L).
\end{equation} 
$\{x_n\}$ ($n=1,\dots,N$) will be the data and the only input that we feed to computer. 
In our analysis below, in most cases we will treat all $N$ data with different temperatures together. In some cases that we specify as temperature resolved analysis, we treat each $N_0$ set of data with a given temperature individually.

The principle component analysis (PCA) method is to find out one or a few directions denoted by $u$, and the project of $(x_n-\bar{x})$ to $u$ can maximally distinguish all data set, where $\bar{x}=1/N\sum_{n} x_n$ is the average of the data set \cite{PRML}. This is equivalent to searching for a vector $u$ such that the variance $\sigma^2$ defined as 
\begin{equation}
\sigma^2=\frac{1}{N}\sum\limits_{n}[u^T(x_n-\bar{x})]^2
\end{equation}
will be maximized. This is further equivalent to finding out the eigen-vector corresponding to the largest or the largest few eigenvalues of the matrix $\mathcal{S}$, which is defined as
\begin{equation}
\mathcal{S}=\frac{1}{N}\sum\limits_{n}(x_n-\bar{x})(x_n-\bar{x})^T.
\end{equation}
$\mathcal{S}$ is a $2L\times 2L$ dimensional matrix in this case and it should have $2L$ eigenvalues, whose eigen-vectors $u_i$ ($i=1,\dots,2L$) form a complete set of bases. If there exist $W$-eigenvalues $\lambda_{k}$ ($k=1,\dots,W$) that are much larger than the rest $\lambda_k$ ($k=W+1,\dots,2L$), and their corresponding eigen-vectors $u_k$ ($k=1,\dots,W)$ are called the ``principle component" and they form the ``principle sub-space". In this case, each vector $x_n$ is characterized by its projection into the principle sub-space, described by a $W$-dimension number $l_n\equiv(l^k_n)=(u^T_k x_n), (k=1,\dots,W)$. In another word, each data is now replaced by 
\begin{equation}
x_n\approx\sum_{k=1}^{W}l_n^k u_k+\sum_{k=W+1}^{2L}(u^T_k \bar{x})u_k.
\end{equation}
Below we will apply this method to $XY$ model in square, triangular and union jack lattices, respectively. We will analyze the outputs of the PCA algorithm, that are, the largest few $\lambda_{k}$, and their corresponding $u_k$, and the structure of $l_n$. We will show that these results are consistent with our understanding of the order parameters in different lattices. 

\begin{figure}[t]
\includegraphics[width=3.3 in]
{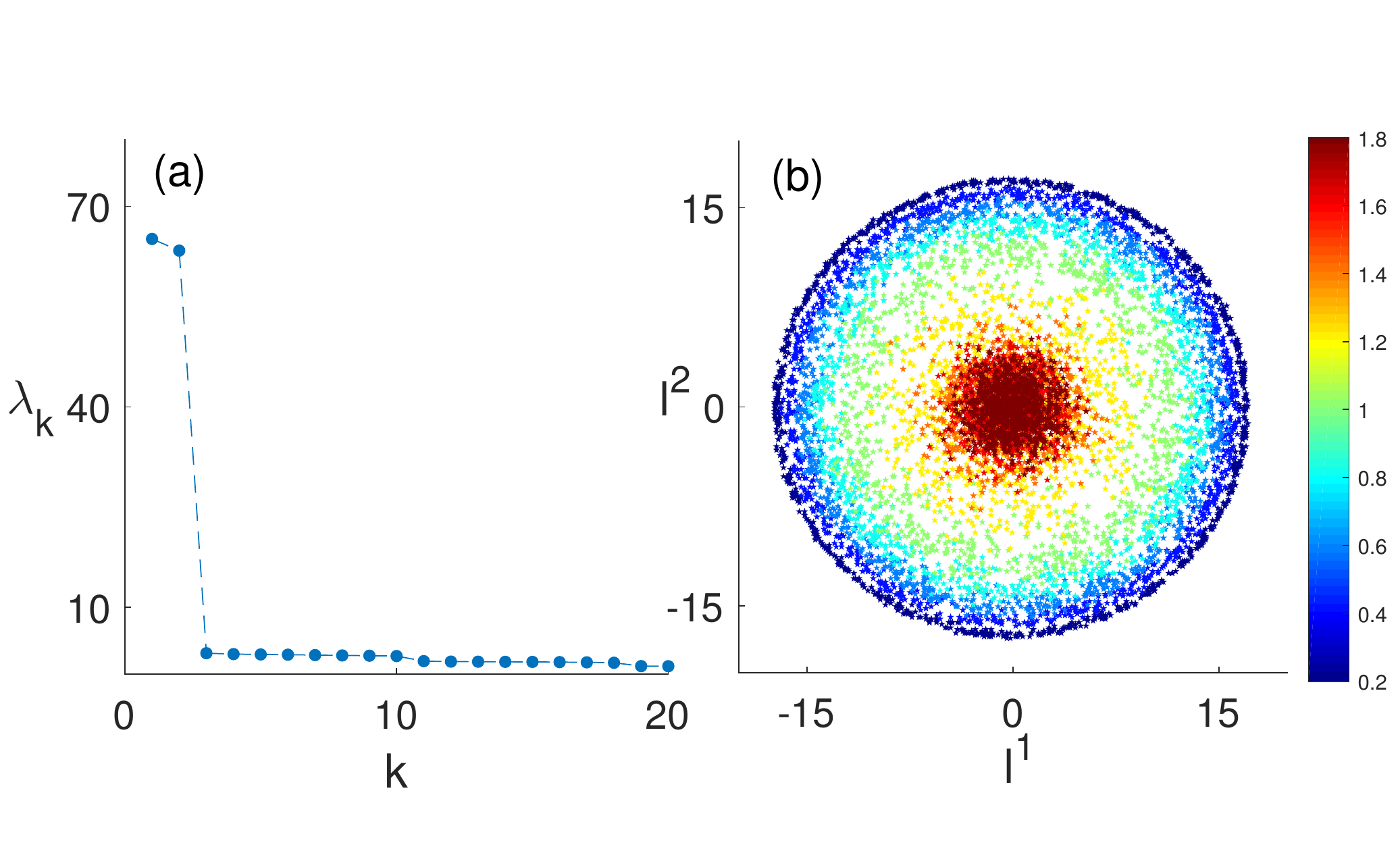}
\caption{(a) The eigen-values of $\mathcal{S}$ matrix for $XY$ model in a two-dimensional $18\times 18$ square lattice; Two eigenvalues are significantly larger than the others. (b) The projection all $N=9000$ data into the two-dimensional principle sub-space $l_n=(l_n^1,l_n^2)$ ($n=1,...,N$). The temperature of the data set ranges from $0.2J$ to $1.8J$, with $\Delta T=0.2J$ and at each temperature $1000$ data sets are taken from classical Monte Carlo simulation. The color bar indicates the temperature at which the datas are generated. }
\label{sq}
\end{figure}

\section{Square Lattices \label{square}}

The PCA analysis for square lattice $XY$ model has been reported in Ref\cite{SqXY_PCA}. Here we review the results and analyze the results based on a simple ``toy model". This viewpoint will be useful for later discussion of triangular and union jack lattices, and the results of the square lattice also help us understand that of the union jack lattice. The results from square lattice are shown in Fig. \ref{sq}. Clearly, in Fig. \ref{sq}(a) one can see that there are two eigen-values that are much larger than all the rest, and their corresponding eigen-vectors are denoted by $u_1$ and $u_2$. Projecting all $x_n$ to $(u_1,u_2)$, we plot all $l_n=(l_n^1,l_n^2)$ in Fig. \ref{sq}(b). One can see that some datas are concentrated nearby the center, and actually they are datas generated at high temperatures; and other datas form a circle and they are generated at low temperatures. The question is that whether and how this outputs are consistent with our prior knowledge of the $XY$ model.  

\begin{figure}[t]
\includegraphics[width=3.3 in]
{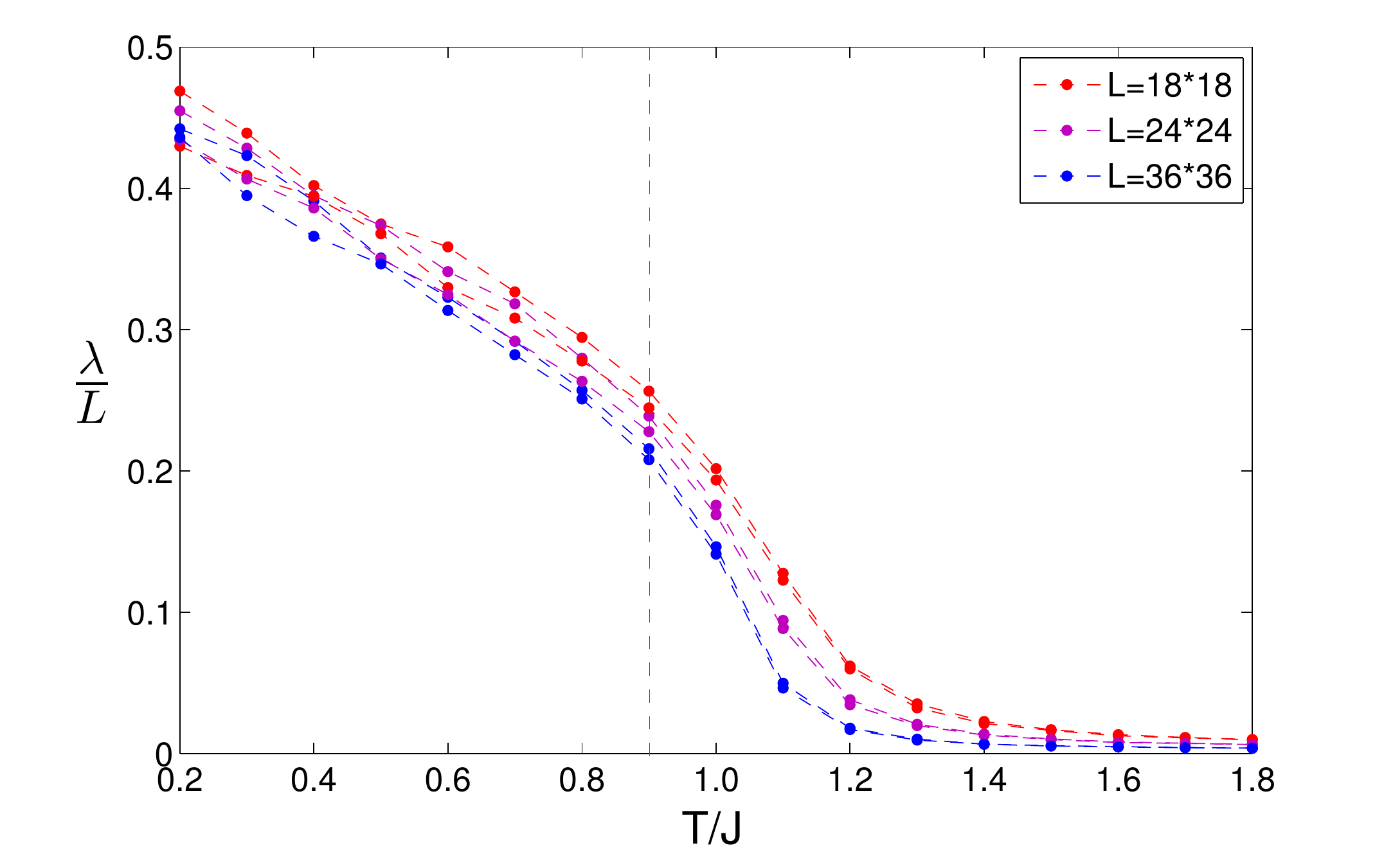}
\caption{Temperature resolved PCA analysis for square lattice $XY$ model in $L=18\times18$, $L=24\times 24$ and $L=36\times 36$ lattices, respectively. We consider temperature range from $0.2J$ to $1.8J$, with $\Delta T=0.1J$ and at each fixed temperature, $N_0=1000$ data set is input. The dashed line indicates the expected temperature for the $KT$ transition in the square lattice $XY$ model.}
\label{Tc_square}
\end{figure}

The insight for understanding this output can be obtained by considering following ``toy model". Let us consider just two sites denoted by $A$ and $B$, and the data will be 
\begin{equation}
x_n=(\cos\theta_A,\cos\theta_B,\sin\theta_A,\sin\theta_B).
\end{equation}
For simplicity, we assume that among all $N$ data set, $p$ percent data mimics the lower temperature case where an antiferromagnetic order is formed, therefore we set $\theta_B=\theta_A+\pi$, with $\theta_A$ uniformly distributing among $\in(0,2\pi]$; and the other $1-p$ percent data mimics higher temperature case where $\theta_A$ and $\theta_B$ are uncorrelated and they are uniformed distributed among $\in(0,2\pi]$. Hence, it is straightforward to obtain, averaging over sufficiently larger number of data set,   
\begin{equation}
\mathcal{S}=(1-p) \mathcal{S}_h+p\mathcal{S}_l,
\end{equation}
where $\mathcal{S}_h=\mathcal{I}/2$, and 
\begin{equation}
\mathcal{S}_l=\left(\begin{array}{cc}\Lambda & 0 \\0 & \Lambda\end{array}\right),
\end{equation}
where 
\begin{equation}
\Lambda=\frac{1}{2}\left(\begin{array}{cc}1 & -1 \\-1 & 1\end{array}\right).
\end{equation}

$\mathcal{S}$ has two degenerate larger eigen-values $\lambda_1=\lambda_2=(1+p)/2$, and the other two eigen-values are $(1-p)/2$. The eigen-vectors corresponding to $\lambda_1$ and $\lambda_2$ are 
\begin{align}
u_1 \propto (1,-1,0,0),\\
u_2 \propto (0,0,1,-1).
\end{align}  
Projecting higher temperature data into $(u_1,u_2)$ sub-space results in $l_n\propto (\cos\theta_A-\cos\theta_B,\sin\theta_A-\sin\theta_B)$, and projecting lower temperature data into $(u_1,u_2)$ sub-space results in $l_n\propto (\cos\theta_A,\sin\theta_A)$.

\begin{figure}[t]
\includegraphics[width=3.0 in]
{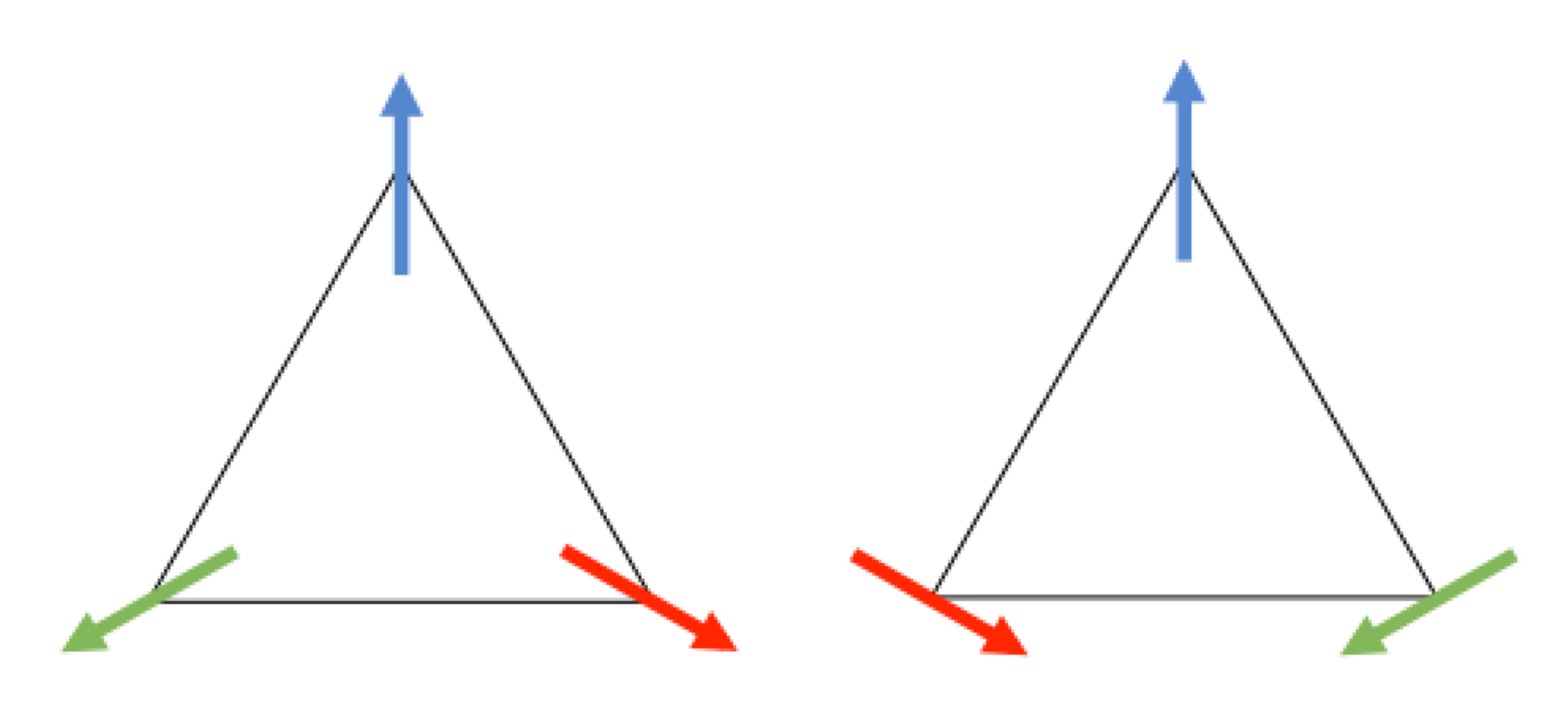}
\caption{Schematic of two types of chiral orders for $XY$ model in triangular lattice.}
\label{tri_chiral}
\end{figure}

This ``toy model" can be straightforwardly generated to a full lattice model by considering $A$ and $B$ as a unit cell and using the translational symmetry to generate a full two-dimensional square lattice. We still consider two types of data. $p$ percent of data simulate the low-temperature case with anti-ferromagneic long rang order, where all $A$ sublattice sites have the same spin angle $\theta$ and the $B$ sublattice sites have spin angle $\theta+\pi$; and the other $1-p$ percent data simulate the high-temperature case where the spin angles at all sites are not correlated at all. We note that this is still an over-simplified situation comparing to real data, but we can obtain some insights from here, because it is straightforward to construct the $\mathcal{S}$ matrix and find out all eigen-values analytically. Interestingly, we find in this case there will be two degenerate larger eigenvalues $(1+(L/2-1)p)/2$, which is much larger than the rest $(1-p)/2$ for large $L$. The eigen-vector corresponding to the larger two eigen-values are
\begin{align}
u_1 \propto (1,-1,\dots, 1,-1,0,0,\dots,0,0),\\
u_2 \propto (0,0,\dots,0,0,1,-1,\dots, 1,-1).
\end{align}  
Projecting higher temperature data into this principle sub-space, and when the system is large enough, $\cos\theta$ or $\sin\theta$ at different sites will average out each other and eventually $l_n \approx (0,0)$. Projecting lower temperature data into this principle sub-space results in $l_n\propto (\cos\theta,\sin\theta)$, which form a circle. This basically shows that the main features in the outputs of the PCA algorithm shown in Fig. \ref{sq}(b) capture the difference between with or without $U(1)$ spin order in lower and higher temperature phases.  
 
From above analysis, one can also see that if all the data are collected from high temperature, that is to say, when $p$ is set to zero, one can see there will be no major eigen-values that are much larger than others. Thus, we perform a temperature resolved PCA analysis, in which all data fed to computer are generated at the same temperature, and then we can plot the normalized eigen-values $\lambda_{1}/L$ and $\lambda_{2}/L$ as a function of temperature of the input data. The result for square lattice is shown in Fig. \ref{Tc_square}. On the other hand, our knowledge of statistical physical already tells us that this model displays a Kosterlitz-Thouless phase transition at $T_c \approx 0.9J$ \cite{XY_Tc}, below which a quasi-long range anti-ferromagnetic order is formed. Indeed, it is found that the two major normalized eigen-values are insensitive to system size for temperature below $T_c$, and for temperature above $T_c$, the two major normalized eigen-values fast decrease as the system size increases.  

\begin{figure}[t]
\includegraphics[width=3.0 in]
{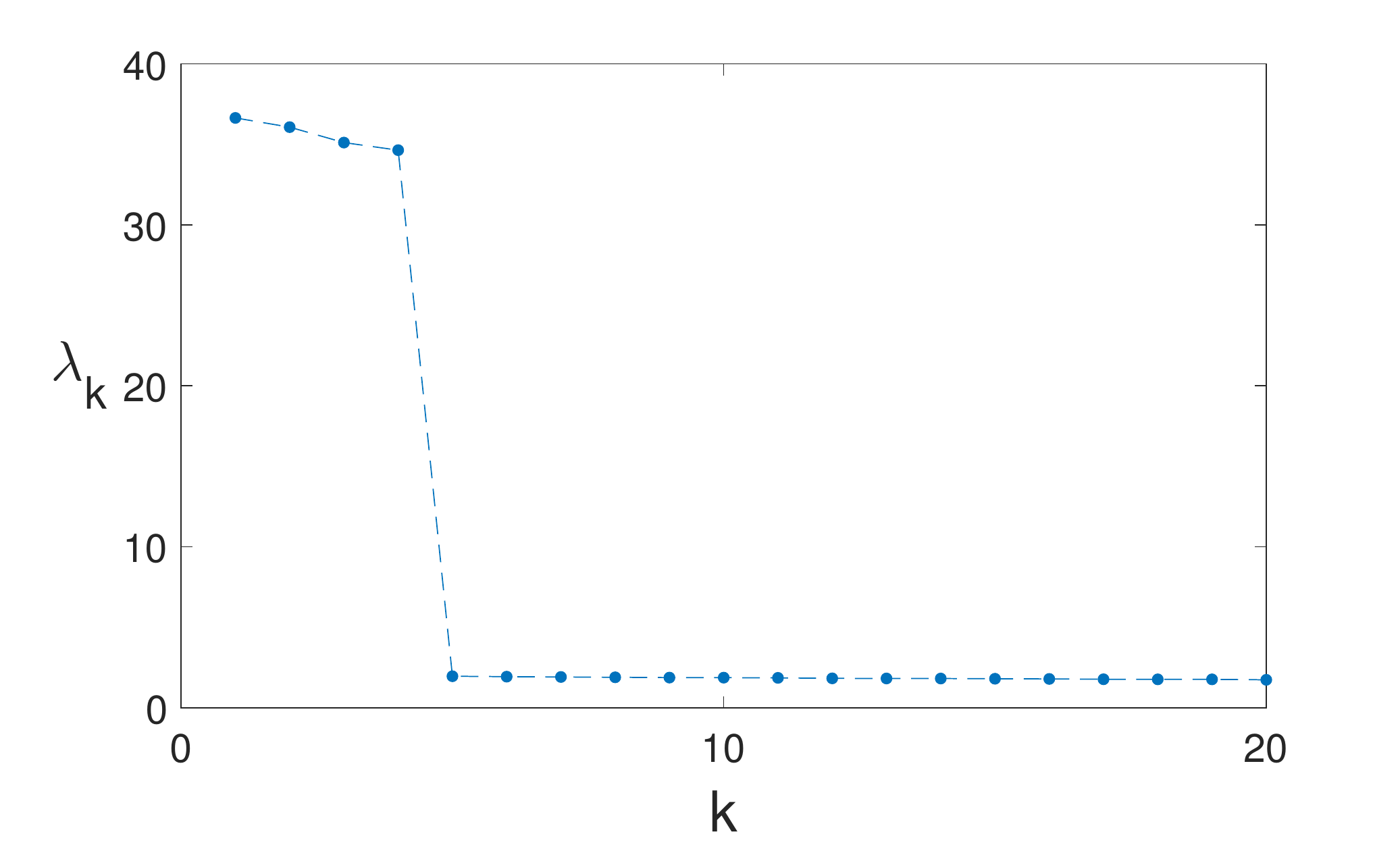}
\caption{The eigen-values of $\mathcal{S}$ matrix for $XY$ model in a two-dimensional $18\times 18$ triangular lattice. Four eigenvalues are significantly larger than the others. The temperature of the data set ranges from $0.3J$ to $0.7J$, with $\Delta T=0.05J$ and at each temperature $N_0=1000$ data set is taken from classical Monte Carlo simulation.}
\label{triangular_eigen}
\end{figure}

\section{Triangular Lattices \label{tri}}

Now we consider $XY$ model in a triangular lattice. For triangular lattice, we know that there are two different order parameters at low-temperatures. One of them is the $U(1)$ spin order whose transition is a Kosterlize-Thouless transition expected at $T_{c1}=0.504J$ \cite{Monte_tri,Monte_tri_uj}. The other order is a $Z_2$ local chiral order defined at each triangle. This chiral order is shown in Fig. \ref{tri_chiral}. The planar spins have a $2\pi/3$ angle between two neighboring sites, and at each triangle, the planar spins rotate either counter-wise or anti-counter-wise, as shown in Fig. \ref{tri_chiral}. This $Z_2$ transition takes place at $T_{c2}=0.512J$ \cite{Monte_tri,Monte_tri_uj} . Since these two transitions are very close in this model, and we do not have enough samples at the temperature window between these two transitions, therefore, the purpose of this model is to discuss the situation with two order parameters but practically it is considered as a single transition. In order to discuss the situation with two separated transitions, for triangular lattice we need huge number of data such that there will be enough samples between two transitions. Instead, this physics is more easy to elaborate with the union jack model in which the two transition temperatures are quite different.   

\begin{figure}[t]
\includegraphics[width=3.5in]
{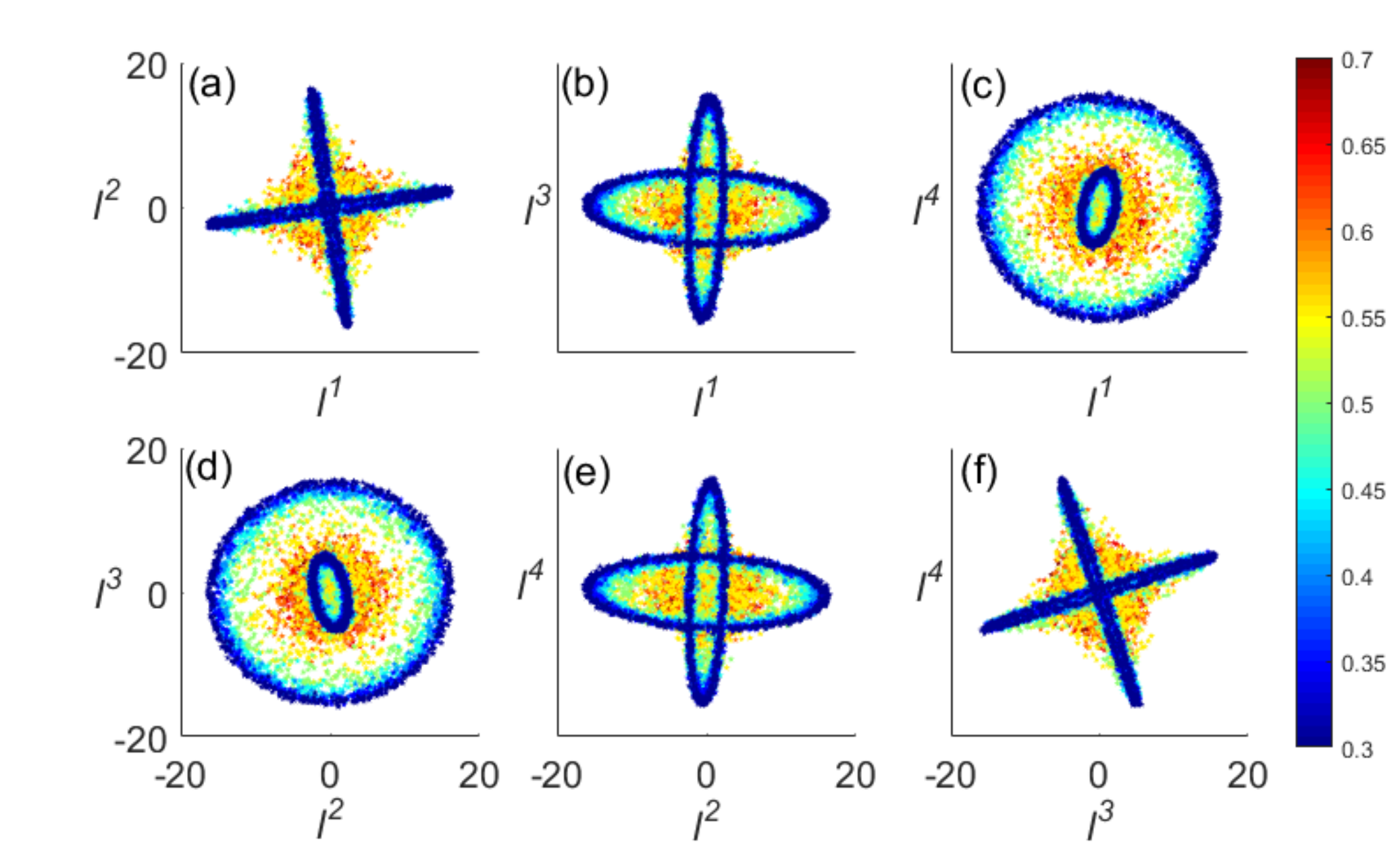}
\caption{The projection of data for $XY$ model in triangular lattice to the principle sub-space. The principle sub-space is generated by data from all temperature. (a-f)  $(l^1_n,l^2_n)$; $(l^1_n,l^3_n)$; $(l^1_n,l^4_n)$; $(l^2_n,l^3_n)$; $(l^2_n,l^4_n)$ and  $(l^3_n,l^4_n)$, respectively. The color bar indicates the temperature at which the datas are generated.}
\label{triangular_projection}
\end{figure}

Similar PCA analysis is performed for data collected from Monte Carlo simulation of nine different temperatures. In Fig. \ref{triangular_eigen} it shows that there are four eigen-values that are much larger than the others. They are denoted by $\lambda_1,\dots,\lambda_4$ and their corresponding eigen-vectors are $u_1,\dots,u_4$, which form the principle sub-space. Thus, each data $x_n$ is now characterized by a four-dimensional number $l_n=(l_n^1,l_n^2,l_n^3,l_n^4)$.  In Fig. \ref{triangular_projection}  we present these numbers for all data.

To understand these results, we perform similar ``toy model" analysis. Now let us consider three sites denoted by $A$, $B$ and $C$. Among all $N$, in $p/2$ percent data, the planar spins rotate clock-wise and the spin angles at three sites are $\theta$, $\theta+2\pi/3$ and $\theta+4\pi/3$, respectively, and for another $p/2$ percent data, the planar spins rotate anti-clock-wise and the spin angles at three sites are $\theta$, $\theta-2\pi/3$ and $\theta-4\pi/3$, respectively, which mimics two different types of chiral orders as shown in Fig. \ref{tri_chiral}. In the rest $1-p$ percent data, the spin angles at three different sites are not correlated. In this case, it is straightforward to calculate that
\begin{equation}
\mathcal{S}=(1-p) \mathcal{S}_h+p\mathcal{S}_l,
\end{equation}
where $\mathcal{S}_h=\mathcal{I}/2$, and 
\begin{equation}
\mathcal{S}_l=\left(\begin{array}{cc}\Lambda & 0 \\0 & \Lambda\end{array}\right),
\end{equation}
and 
\begin{equation}
\Lambda=\frac{1}{2}\left(\begin{array}{ccc}1 & -\frac{1}{2} & -\frac{1}{2} \\-\frac{1}{2} & 1 & -\frac{1}{2} \\ -\frac{1}{2} & -\frac{1}{2}& 1\end{array}\right).
\end{equation}

\begin{figure}[t]
\includegraphics[width=3.6 in]
{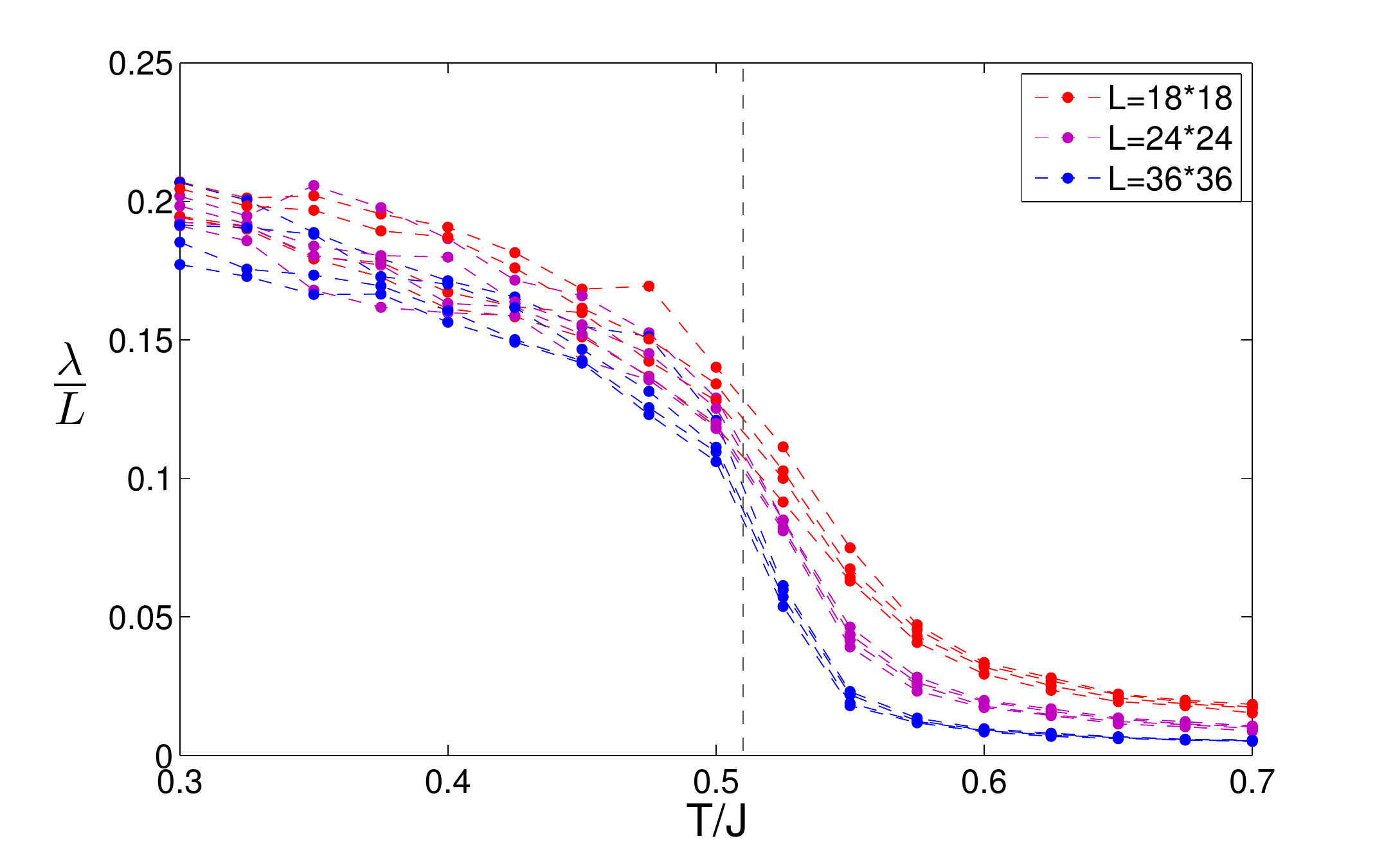}
\caption{Temperature resolved PCA analysis for triangular lattice $XY$ model in $L=18\times18$, $L=24\times 24$ and $L=36\times 36$ lattices, respectively. We consider temperature range from $0.3J$ to $0.7J$, with $\Delta T=0.025J
$ and at each fixed temperature, $N_0=1000$ data set is input. The dashed line indicates the expected temperature for the $KT$ transition in the triangular lattice $XY$ model.}
\label{Tc_triangle}
\end{figure}

Note that two eigen-values of $\mathcal{S}_l$ are $3/4$ and the other is zero, the corresponding eigen-vector for eigen-value $3/4$ can be written as
\begin{align}
&d_{1}\propto \left(1, \cos\left(\frac{2\pi}{3}\right), \cos\left(\frac{4\pi}{3}\right)\right)\propto (2,-1,-1),\\
&d_{2}\propto \left(0, \sin\left(\frac{2\pi}{3}\right), \sin\left(\frac{4\pi}{3}\right)\right)\propto (0,1,-1).
\end{align}
Therefore, the matrix $\mathcal{S}$ has four degenerate eigen-value as $1/2+p/4$ and the rest two as $1/2-p/2$. Corresponding to the four degenerate eigen-values, one way to choose the orthogonal eigen-vectors are
\begin{align}
&u_1=(d_1,d_2); \label{u1}\\
&u_2=(d_2,d_1); \label{u2}\\
&u_3=(d_1,-d_2); \label{u3}\\
&u_4=(-d_2,d_1). \label{u4}
\end{align}
Projecting the low-temperature data into this sub-space, for samples with clock-wise chirality, $l_n=(3\cos\theta,0,0,3\sin\theta)$; and for samples with anti-clock-wise chirality, $l_n=(0,3\sin\theta,3\cos\theta,0)$. Keeping in mind that $\theta$ is uniformly distributed among $(0,2\pi]$, plotting $(l_n^1,l_n^2)$, $(l_n^1,l_n^3)$, $(l_n^2,l_n^4)$ and $(l_n^3,l_n^4)$ for all data give a cross, and the data with different chirality set at different lines of the cross. And when plotting $(l_n^1,l_n^4)$, the clock-wise chirality sample form a circle and the anti-clock-wise chirality samples are concentrated at the center. The situation is reversed when plotting $(l_n^2,l_n^3)$. 

Because all the four eigen-values are degenerate, one can also make a rotation for the bases $u_1,\dots,u_4$. For instance, one can choose 
\begin{align}
&u_1^\prime=\cos\alpha u_1+\sin\alpha u_2; \label{rot1}\\
&u_2^\prime=-\sin\alpha u_1+\cos\alpha u_2; \label{rot2}\\
&u_3^\prime=\cos\beta u_3+\sin\beta u_4; \label{rot3}\\
&u_4^\prime=-\sin\beta u_3+\cos\beta u_4. \label{rot4}
\end{align}
Under this bases, for samples with clock-wise chirality, 
\begin{equation}
l_n=3(\cos\alpha\cos\theta,-\sin\alpha\cos\theta,\sin\beta\sin\theta,\cos\beta\sin\theta),
\end{equation}
and for samples with anti-clock-wise chirality,
\begin{equation}
l_n=3(\sin\alpha\sin\theta,\cos\alpha\sin\theta,\cos\beta\cos\theta,-\sin\beta\cos\theta).
\end{equation}
In this case, plotting $(l_n^1,l_n^2)$ and $(l^3_n,l^4_n)$ give rise to two tilted cross, as shown in the low temperature datas (deep blue) in (a) and (f) of Fig. \ref{triangular_projection}. Without loss of generality, we can assume both $\alpha$ and $\beta$ belong to $(0,\pi/4]$,  plotting $(l_n^1,l_n^3)$ and $(l^2_n,l^4_n)$ results in two ellipses, with the long axes of one ellipse is along $l^1$ and the other is along $l^2$, as (b) and (e) of Fig. \ref{triangular_projection}. Plotting $(l_n^1,l_n^4)$ and $(l^2_n,l^3_n)$ also results in two ellipses, with one is always larger than the other, as (c) and (d) of Fig. \ref{triangular_projection}.

\begin{figure}[t]
\includegraphics[width=3.3 in]
{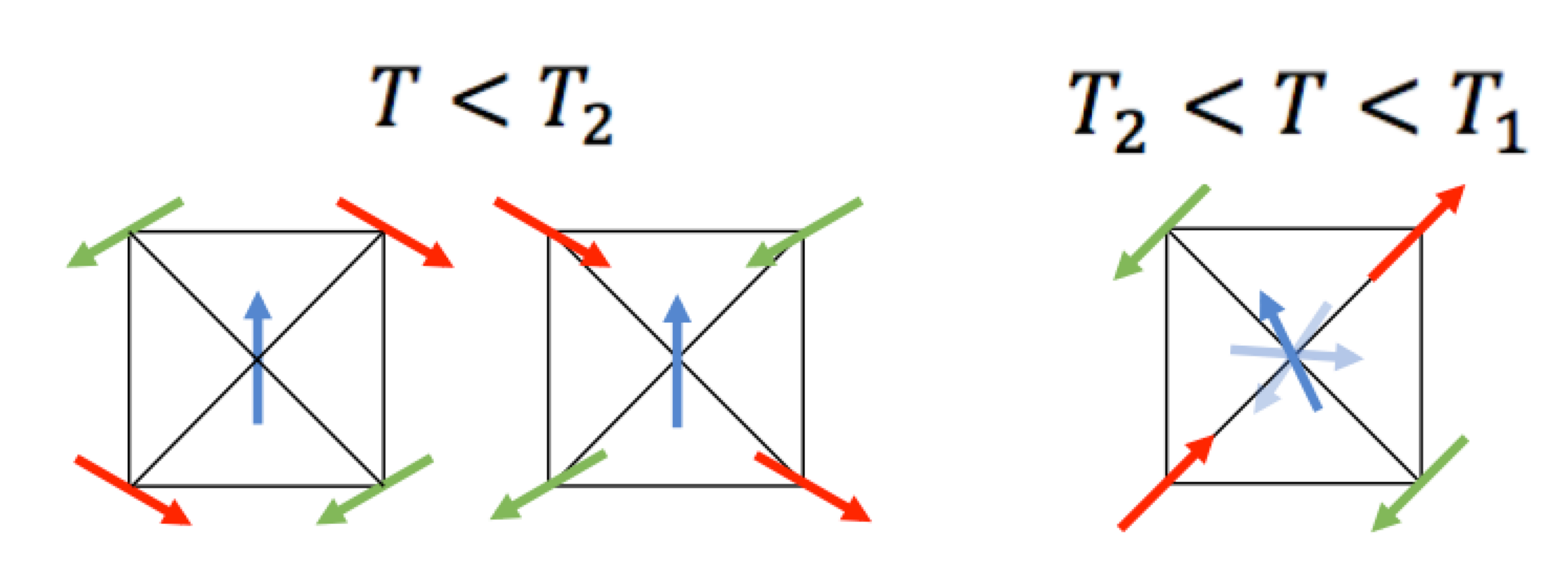}
\caption{Schematic of two types of chiral orders for $XY$ model in the union jack lattice in low-temperature, and an intermediate anti-ferromagnetic order at intermediate temperature.}
\label{UJ_AF}
\end{figure}

Here we shall also remark that the appearance of $l_n$ is quite sensitive to the choice of bases. For instance, instead of using Eq. \ref{u1}-\ref{u4}, one can also choose the bases as 
\begin{align}
&u_1=(d_2,0); \label{u11}\\
&u_2=(0,d_2); \label{u21}\\
&u_3=(d_1,0); \label{u31}\\
&u_4=(0,d_1). \label{u41}
\end{align}
Under this bases, $l_n=\frac{3}{2}(-\sin\theta,\cos\theta,\cos\theta,\sin\theta)$ for low-temperature data with clock-wise chirality, and $l_n=\frac{3}{2}(\sin\theta,-\cos\theta,\cos\theta,\sin\theta)$ for low-temperature data with anti-clock-wise chirality. In this case, $(l^1_n,l^2_n)$, $(l^1_n,l^3_n)$, $(l^2_n,l^4_n)$ and $(l^3_n,l^4_n)$ are all circles, and $(l^1_n,l^4_n)$, $(l^2_n,l^3_n)$ are two crosses at $45$-degree. Furthermore, one can rotate the bases with the same transformation as Eq. \ref{rot1}-\ref{rot4}, and under these rotated bases, for clock-wise and anti-clock-wise samples, $l_n=\frac{3}{2}(-\sin(\theta-\alpha),\cos(\theta-\alpha),\cos(\theta+\beta),\sin(\theta+\beta))$, and $l_n=\frac{3}{2}(\sin(\theta-\alpha),-\cos(\theta-\alpha),\cos(\theta+\beta),\sin(\theta+\beta))$, respectively. In this case, $(l^1_n,l^2_n)$ and $(l^3_n,l^4_n)$ will remain as two circles, and other four combinations of $(l^1_n,l^3_n)$, $(l^1_n,l^4_n)$, $(l^2_n,l^3_n)$ and $(l^2_n,l^4_n)$ will all become two ellipses perpendicular to each other and orientated at $45$- and $135$-degrees. We will come back to this point in the discussion of the union jack lattice. 

In this ``toy model" with only three sites, the eigen-values of the four degenerate ones are not significantly larger than the rest two, however, similar as sec. \ref{square} one can consider these three sites as a unit cell and generate a triangular lattice model. Similarly, we assume that in all data, the spin angles either are identical for all unit cells, simulating the low-temperature situation, or they are uncorrelated at all, simulating the high-temperature situation. In this case, one can show that the four degenerate eigen-values become $1/2+p(L-2)/4$, and these four are much larger than the rest which remain as $(1-p)/2$. Similarly, when projecting the high temperature date into the principle sub-space, the contribution from different sites also average out and one obtains $l_n\approx (0,0,0,0)$, as one can also see in Fig. \ref{triangular_projection}. In this way, we understand the output of the principle sub-space and the projection of the date from both ordered and disordered phase into this sub-space.  

\begin{figure}[t]
\includegraphics[width=3.3 in]
{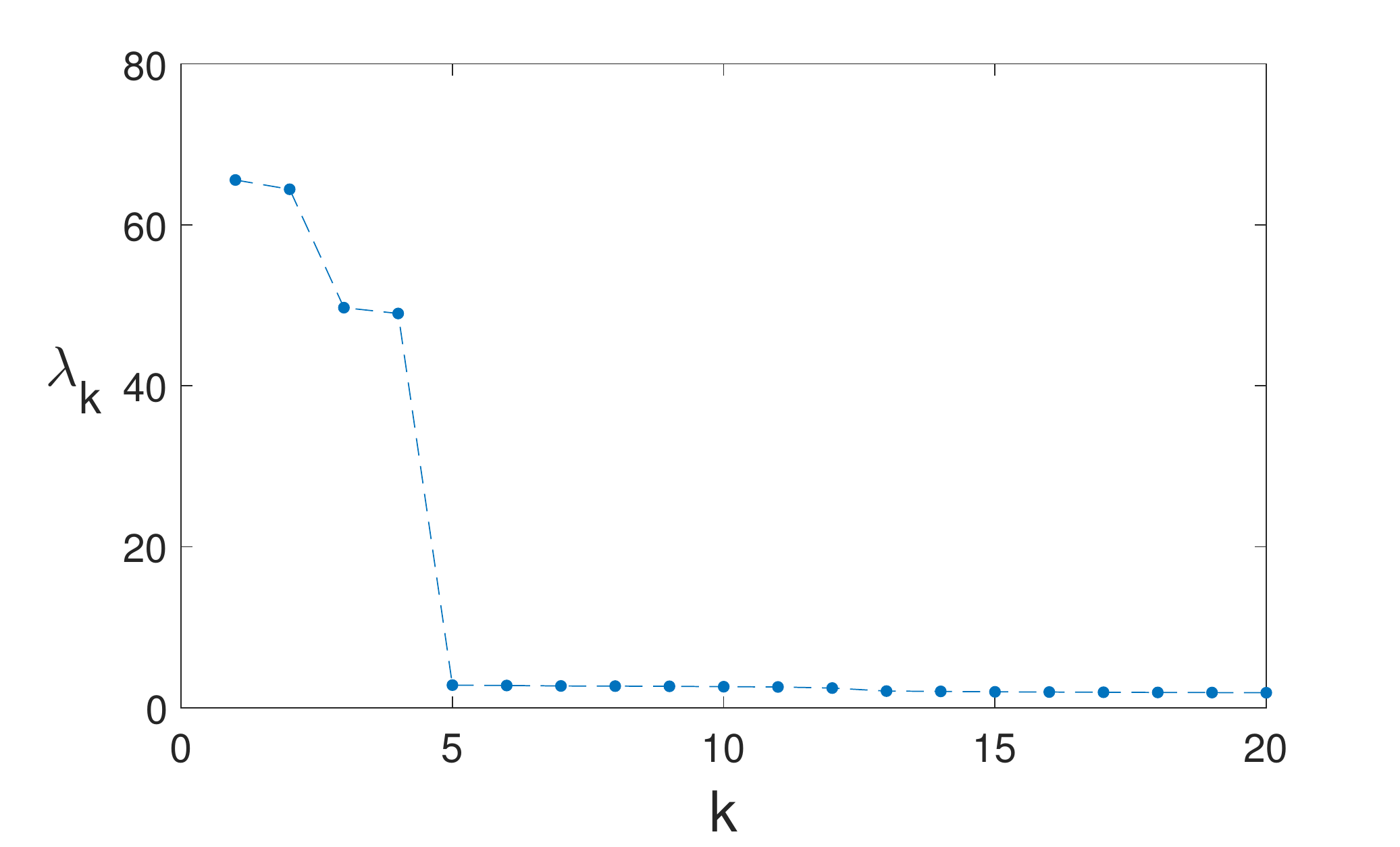}
\caption{The eigen-values of $\mathcal{S}$ matrix for $XY$ model in a two-dimensional $18 \times 18$ union jack lattice. Four eigenvalues are significantly larger than the others. The temperature of the data set ranges from $0.2J$ to $1.0J$, with $\Delta T=0.1J$ and at each temperature $N_0=1000$ data set is taken from classical Monte Carlo simulation.}
\label{UJ_eigenvalue}
\end{figure}

Before concluding this session, we also show in Fig. \ref{Tc_triangle} a temperature resolved PCA analysis that reveals how the large principle component depends on temperature. Similar as the square lattice case, below certain temperature the four normalized major principle eigenvalues are not sensitive to system size, and above certain temperature, they decrease to quite small ones as the system size increases. This transition temperature scale is consistent with the KT transition expected for $XY$ model in a triangular lattice. 

\section{Union Jack Lattice}

Now we move to discuss the $XY$ model in union jack lattice\cite{Monte_tri_uj}. The union jack lattice is made of a square lattice and an extra site at each center of the plaquette. The physics of $XY$ model in union jack lattice is shown in Fig. \ref{UJ_AF}. For $T>T_1$, it is a spin disordered normal phase. Between $T_1>T>T_2$, the $\pi$-antiferromagnetic spin order forms in the square lattice, as shown in Fig. \ref{UJ_AF}(b). Further lowering temperature to $T<T_2$, a $2\pi/3$-antiferromagntic order is formed at each triangle, as shown in Fig. \ref{UJ_AF}(a). Similar to the antiferromagnetic order in triangular lattice, it has two opposite chirality. The phase transition at $T_1$ is a Kosterlitz-Thouless type of $U(1)$ ordered transition, and the transition at $T_2$ is an Ising transition. 

\begin{figure}[t]
\includegraphics[width=3.5 in]
{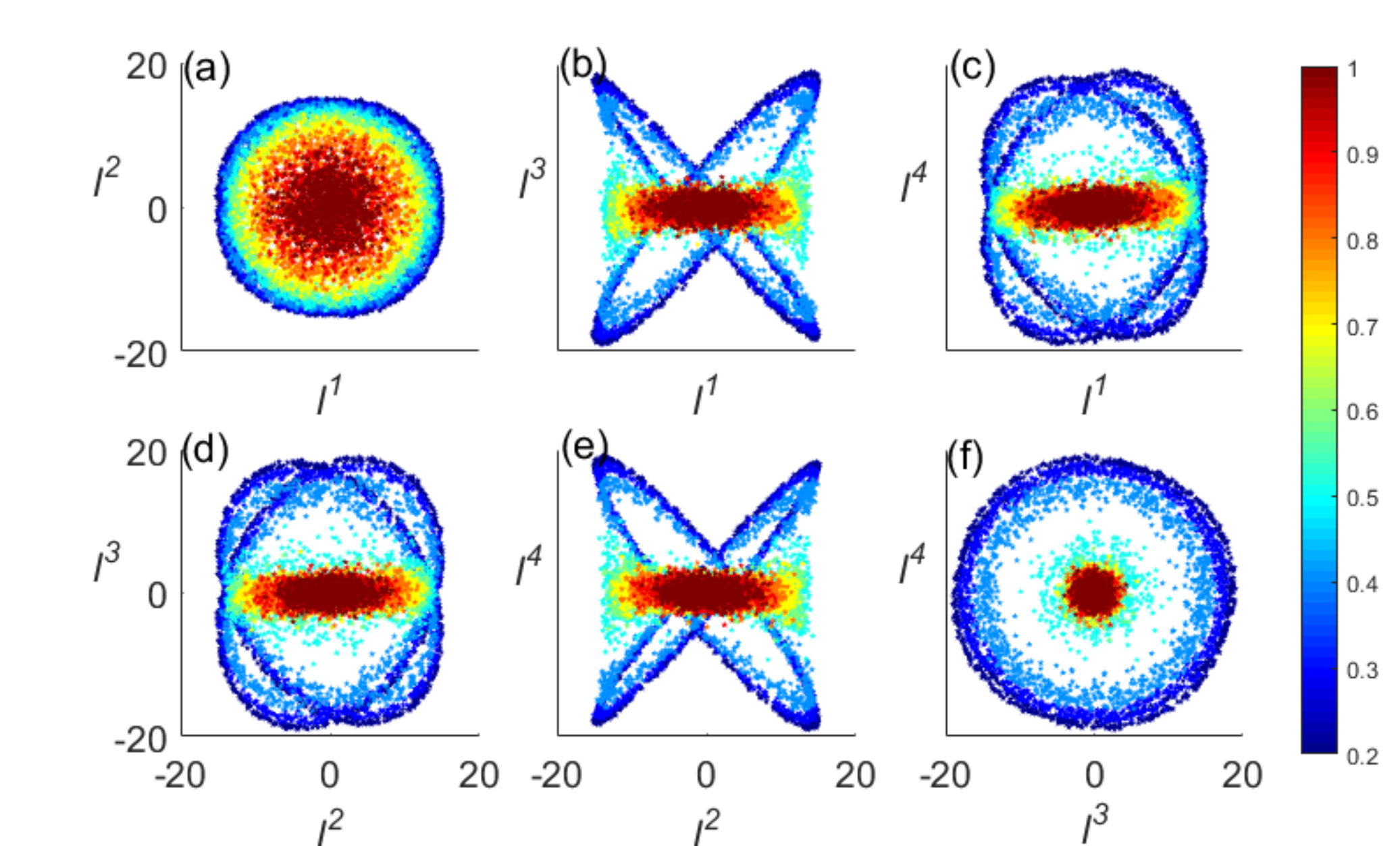}
\caption{The projection of data for $XY$ model in union jack lattice to the principle sub-space. (a-f) correspond to $(l^1_n,l^2_n)$; $(l^1_n,l^3_n)$; $(l^1_n,l^4_n)$; $(l^2_n,l^3_n)$; $(l^2_n,l^4_n)$ and $(l^3_n,l^4_n)$, respectively. The color bar indicates the temperatures at which the datas are generated.}
\label{UJ_projection}
\end{figure}

In the union jack lattice, $T_1=0.64J$ and $T_2=0.43J$ \cite{Monte_tri_uj}. Since these two transition temperatures are well separated, we will have enough data sampled at all three temperature regimes, i.e. $T>T_1$, $T_2<T<T_1$ and $T<T_2$. Therefore we will use this model to address the issue of two phase transitions. We perform PCA analysis for date collected from both three temperature regimes. The results are shown in Fig. \ref{UJ_eigenvalue} and Fig. \ref{UJ_projection}. In Fig. \ref{UJ_eigenvalue} we can see that there are four eigenvalues that are considerably larger the others. The corresponding four eigenvectors form the principle sub-space. The projection of all date into this principle sub-space is shown in Fig. \ref{UJ_projection}.

In this case, the low-temperature ordered phase has four cites at each unit cell. Similar to the discussion in above two sessions, we can design a ``toy model" with four sites to understand this behavior. Here we will not repeat the similar analysis here. In fact, for temperature below $T_2$, the situation is nearly the same as the low temperature phase of the triangular lattice. Using the bases rotation of Eq. \ref{u11}-\ref{u41}, as we have discussed in the sec. \ref{tri}, two combinations $(l^1_n,l^2_n)$ and $(l^3_n,l^4_n)$ behaves as two circles, and the other four combinations of $(l^1_n,l^3_n)$, $(l^1_n,l^4_n)$, $(l^2_n,l^3_n)$ and $(l^2_n,l^4_n)$ are two ellipses perpendicular to each other. In fact, as one can see from Fig. \ref{UJ_projection} the data generated below $T_2$ behave in this way.

The reason we use the bases of Eq. \ref{u11}-\ref{u41} to discuss the union jack lattice worth emphasizing. It is an important observation that when there is only a $\pi$-anti-ferromagnetic order in the square lattice as in the intermediate temperature, as we discussed in sec. \ref{square}, there are two major components and their eigen-values are consistent with that of $d_2$. Hence, for data generated at the intermediate temperature, $(l^3_n,l^4_n)\approx (0,0)$ because the site at the center of the plaquette is not ordered, $(l^1_n,l^2_n)$ forms a circle as the square lattice case, and $(l^1_n,l^3_n)$, $(l^1_n,l^4_n)$, $(l^2_n,l^3_n)$ and $(l^1_n,l^4_n)$ all behave as a one-dimensional line, because it is a projection of the circle onto one of the axes. This is exactly what are found in Fig. \ref{UJ_projection}.

\begin{figure}[t]
\includegraphics[width=3.5 in]
{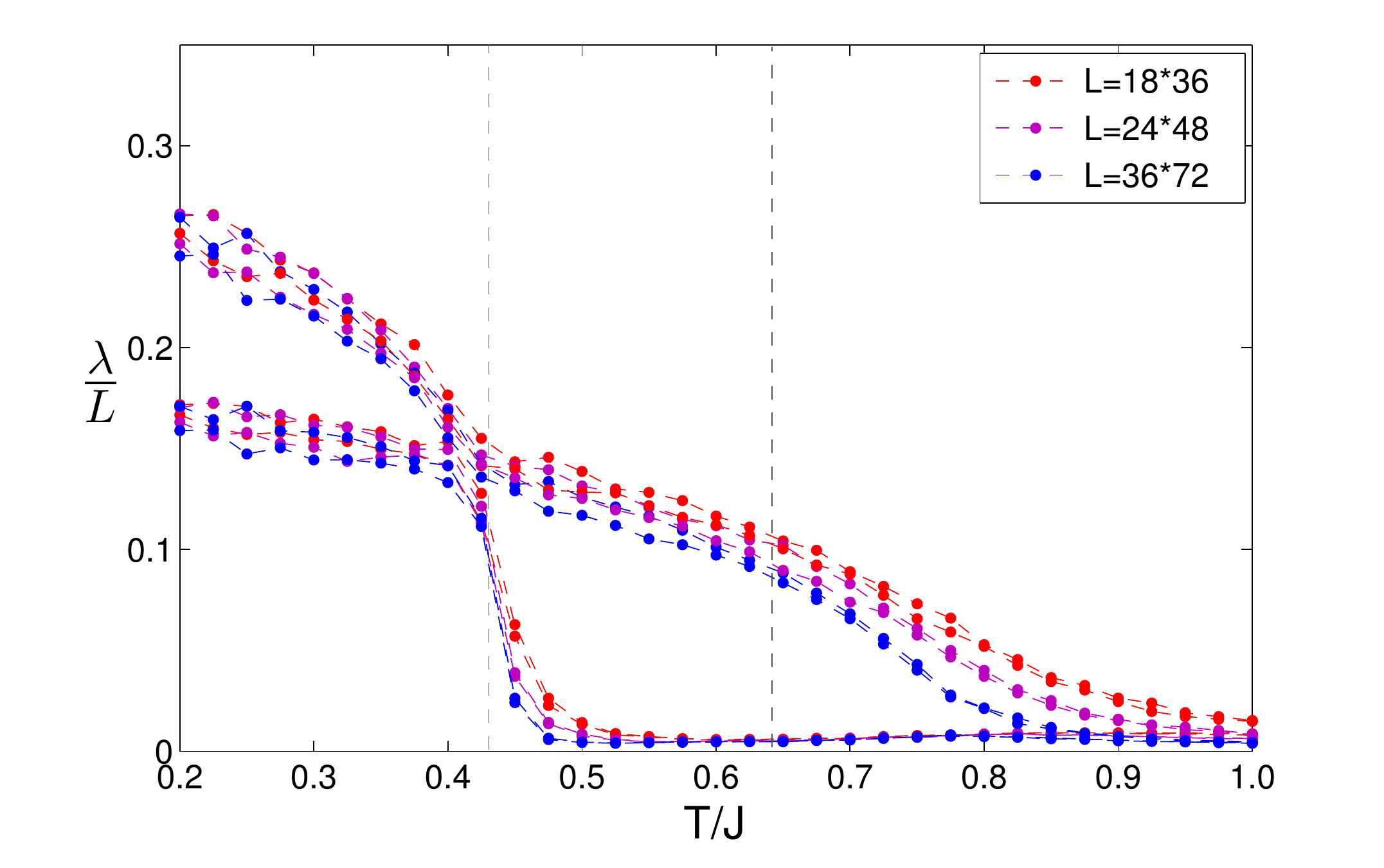}
\caption{Temperature resolved PCA analysis for union jack lattice $XY$ model in $L=18\times 18$, $L=24\times 24$ and $L=36\times 36$ lattices, respectively. We consider temperature range from $0.2J$ to $1.0J$, with $\Delta T=0.025J$ and at each fixed temperature, $N_0=1000$ data set is input. The dashed line indicates the expected temperature for the Ising transition and  the $KT$ transition in the union jack lattice $XY$ model.}
\label{Tc_UJ}
\end{figure}

In Fig. \ref{Tc_UJ}, we perform a temperature resolved PCA analysis. It is very clear that there are four major eigen-values at the lowest temperature. Then, two of them vanish at a temperature $\sim 0.43J=T_2$, and the other two gradually vanish as temperature increases. Similar as the square and the triangular lattice cases, the other two eigen-values are not sensitive to system size below $T_1$, but above $T_1$ they decrease as the system size increases. Here we should stress that the different behavior of these eigen-values at $T_1$ and $T_2$ are in fact quite physical. At $T_2$ the Ising transition is a second-order phase transition, and the eigen-values indeed vanish pretty much close to the expected transition temperature. While $T_1$ is a Kosterlize-Thouless transition and the transition takes place in a much smooth way. 

\section{Outlook}

In this work we utilize PCA analysis to classify the classical Monte Carlo data generated for $XY$ model in a two-dimensional square, triangular and union jack lattices, where the latter two are frustrated lattices and display two order parameters, and the last one even has two well-separated phase transition. Using simple ``toy" models, we show that the outputs of the PCA analysis fully agree with our prior understanding of different orders in these models, and the temperature resolved analysis of the principle components are also consistent with the critical temperature and the order of transitions. 

Although this simple PCA analysis is good enough to recognize different phases, the outputs are still too complicated to directly read out the order parameters. The physical reason is that some order parameters, such as the chirality, is a non-linear function of the input function $\cos\theta$ or $\sin\theta$, and the kernel PCA can directly reveal these non-linear order parameters\cite{NewPaper} . Moreover, to reveal the Kosterlize-Thouless transition more directly, one also needs some non-linear order parameters such as superfluid density \cite{XY1,XY2} and the kernel PCA can also help. Finally we will also use the neural network to find out the best kernel. These results will be published in the following publications.  

\acknowledgments
We thank Youjin Deng for helpful discussion. This work is supported by NSFC Grant No. 11325418 and MOST under Grant No. 2016YFA0301600.


\begin{thebibliography}{99}
\bibitem{Hubbard} A. Montorsi, The Hubbard Model: A Reprint Volume (World Scientific, 1992).
\bibitem{FH_exp} A. Mazurenko, C. S. Chiu, G. Ji, M. F. Parsons, M. Kanász-Nagy, R. Schmidt, F. Grusdt, E. Demler, D. Greif, M. Greiner, Nature {\bf 545}, 462 (2017) 
\bibitem{PRML} C. M. Bishop, Pattern Recognition And Machine Learning (Springer, 2007). 
\bibitem{Ising_PCA} L. Wang, Phys. Rev. B {\bf 94},195105 (2016).
\bibitem{Ising_nn} J. Carrasquilla, R. G.Meiko, Nat. Phy. {\bf 13}, 431 (2017).
\bibitem{Confusion} E. P. L. van Nieuwenburg, Y. H. Liu and S. D. Huber, Nat. Phy. {\bf 13}, 435 (2017).
\bibitem{Ising_BM} G. Torlai and R. G. Melko, Phys. Rev. B {\bf 94}, 165134 (2016)
\bibitem{Ising_XY_VAE} S. Wetzel, arXiv:1703.02435.
\bibitem{Ising_SVM} P. Ponte, R. G. Melko, arXiv:1704.05848.
\bibitem{SqXY_PCA}W. J. Hu, R. Singh, Richard Scalettar, arXiv:1704.00080.
\bibitem{Ising_SU(2)_nn} S. Wetzel, M. Scherzer, arXiv:1705.05582.
\bibitem{CNN_Fermions1} K. Ch'ng, J. Carrasquilla, R. G. Melko and E. Khatami, arXiv:1609.02552.
\bibitem{CNN_Fermions2} P. Broecker, J. Carrasquilla, R. G. Melko and S. Trebst, arXiv:1608.07848.
\bibitem{Liang3} J. Liu, H. Shen, Y. Qi, Z. Y. Meng and L. Fu, Phys. Rev. B {\bf 95}, 241104 (2017)
\bibitem{WangLei1}L. Huang, L. Wang, arXiv: 1610.02746.
\bibitem{WangLei2}L. Huang, Y. F. Yang, L. Wang, arXiv: 1612.01871.
\bibitem{Liang2} X. Y. Xu, Y. Qi, J. Liu, L. Fu and Z. Y. Meng, arXiv: 1612.03804
\bibitem{WangLei3}L. Wang, arXiv:1702.08586.
\bibitem{Liang1} Y. Nagai, H. Shen, Y. Qi, J. Liu and L. Fu, arXiv: 1705.06724


\bibitem{FXY_theory5}J. Villain, J. Phys. C 10, 1717 (1977).
\bibitem{FXY_theory6}J. Villain, J. Phys. C 10, 4793 (1977).
\bibitem{FXY_theory1}D. H. Lee, J. D. Joannopoulos, J. W. Negele, and D. P. Landau, Phys. Rev. Lett. 52, 433 (1984).
\bibitem{FXY_theory2}S. Miyashita and H. Shiba, J. Phys. Soc. Jpn 53, 1145 (1984).
\bibitem{FXY_theory3}S. Lee and K. C. Lee, Phys. Rev. B 49, 15184 (1994).
\bibitem{FXY_theory4}S. Korshunov, Phys. Rev. Lett. 88, 167007 (2002).
\bibitem{FXY_theory0}M. Hasenbusch, A. Pelissetto, and E. Vicari, J. Stat. Mech.: Theory Exp. P12002 (2005) 
\bibitem{Monte_tri}T. Obuchi and H. Kawamura, J. Phys. Soc. Jpn. 81, 054003 (2012).
\bibitem{Monte_tri_uj}J. P. Lv, T. M. Garoni, Y. J. Deng, Phys. Rev. B 87, 024108(2013).
\bibitem{XY_Tc} P. Olsson, Phys. Rev. B {\bf 52} 4526 (1995).
\bibitem{NewPaper} C. Wang and H. Zhai, to appear. 
\bibitem{XY1} T. Ohata and D. Jasnow, Phys. Rev. B 20.139(1978).
\bibitem{XY2} H. Weber and P. Minnhagen, Phys. Rev. B 37.5986(1987).

\end{thebibliography}
\end{document}